\documentclass[12pt]{article}
\usepackage{epsfig}
\usepackage{color}
\usepackage{a4}
\usepackage{latexsym}
\usepackage{cite}

\usepackage{pslatex}
\usepackage[latin1]{inputenc}
\usepackage[T1]{fontenc}

\usepackage{color,colordvi}
\def\colour4colour#1{\Blue{#1}}

\newcommand{\gsim}{\raisebox{-0.07cm}{$\:\:\stackrel{>}{{\scriptstyle
 \sim}}\:\, $} }
\newcommand{\lsim}{\raisebox{-0.07cm}{$\:\:\stackrel{<}{{\scriptstyle
 \sim}}\:\, $} }
\newcommand{\equal}{\:\: = \:\:}

\newcommand{\hspn}{{\hspace{-4mm}}}

\newcommand{\beq}{\begin{equation}}
\newcommand{\eeq}{\end{equation}}
\newcommand{\bea}{\begin{eqnarray}}
\newcommand{\eea}{\end{eqnarray}}
\newcommand{\nn}{\nonumber}
\newcommand{\MSb}{$\overline{\mbox{MS}}$}

\newcommand{\als}{\alpha_{\rm s}}
\newcommand{\ars}{a_{\rm s}}

\begin{document}
\setlength{\parskip}{0.2cm}
\setlength{\baselineskip}{0.52cm}

\def\nc{{n_c}}
\def\ncs{{n_{c}^{\,2}}}
\def\nct{{n_{c}^{\,3}}}
\def\ncf{{n_{c}^{\,4}}}

\def\mus{{\mu_{\! f}^{\,2}}}

\def\z#1{{\zeta_{#1}}}
\def\zss{\zeta_2^{\,2}}

\def\ca{{C^{}_A}}
\def\cas{{C^{\,2}_A}}
\def\cat{{C^{\,3}_A}}
\def\caf{{C^{\,4}_A}}
\def\cf{{C^{}_F}}
\def\cfs{{C^{\, 2}_F}}
\def\cft{{C^{\, 3}_F}}
\def\cff{{C^{\, 4}_F}}
\def\nf{{n^{}_{\! f}}}
\def\nfs{{n^{\,2}_{\! f}}}
\def\nft{{n^{\,3}_{\! f}}}

\def\dfRAnr{{ {d_{\,R}^{\,abcd}\,d_{\,A}^{\,abcd} \over n_c} }}
\def\dfRRnr{{ {d_{\,R}^{\,abcd}\,d_{\,R}^{\,abcd} \over n_c} }}
\def\dabctnr{{ {d^{abc}d_{abc}}\over{n_c} }}

\def\dfAAna{{ {d_{\,A}^{\,abcd}\,d_{\,A}^{\,abcd} \over n_a} }}
\def\dfRAna{{ {d_{\,R}^{\,abcd}\,d_{\,A}^{\,abcd} \over n_a} }}
\def\dfRRna{{ {d_{\,R}^{\,abcd}\,d_{\,R}^{\,abcd} \over n_a} }}

\def\als{{\alpha_{\rm s}}}
\def\as(#1){{\alpha_{\rm s}^{\:#1}}}
\def\ar(#1){{a_{\rm s}^{\:#1}}}

\def\frct#1#2{\mbox{\large{$\frac{#1}{#2}$}}}

\begin{titlepage}
\noindent
%
%
DESY 21-203 \hfill November 2021\\
NIKHEF 21-030\\ 
LTH 1282 \\
\vspace{1.0cm}
\begin{center}
\Large
{\bf Low moments of the four-loop splitting functions in QCD} \\
\vspace{1.5cm}
\large
S. Moch$^{\, a}$, B. Ruijl$^{\, b}$, T. Ueda$^{\, c}$, 
J.A.M. Vermaseren$^{\, d}$ and A. Vogt$^{\, e}$\\
\vspace{1.5cm}
\normalsize
{\it $^a$II.~Institute for Theoretical Physics, Hamburg University\\
Luruper Chaussee 149, D-22761 Hamburg, Germany}\\
\vspace{4mm}
{\it $^b$ETH Z\"urich\\
R\"amistrasse 101, CH-8092 Z\"urich, Switzerland}\\
\vspace{4mm}
{\it $^c$ Department of Materials and Life Science, Seikei University\\
3-3-1 Kichijoji Kitamachi, Musashino-shi, Tokyo 180-8633, Japan}\\
\vspace{4mm}
{\it $^d$Nikhef Theory Group \\
Science Park 105, 1098 XG Amsterdam, The Netherlands} \\
\vspace{4mm}
{\it $^e$Department of Mathematical Sciences, University of Liverpool\\
Liverpool L69 3BX, United Kingdom}\\
\vspace{2.0cm}
%
{\large \bf Abstract}
\vspace{-0.2cm}
\end{center}
We have computed the four lowest even-$N$ moments of all four splitting
functions for the evolution of flavour-singlet parton densities of
hadrons at the fourth order in the strong coupling constant $\als$.
The perturbative expansion of these moments, and hence of the splitting
functions for momentum fractions $x \gsim 0.1$, is found to be well 
behaved with relative $\als$-coefficients of order one and sub-percent 
effects on the scale derivatives of the quark and gluon distributions
at $\als \lsim 0.2$. More intricate computations, including other 
approaches such as the operator-product expansion, are required to cover 
the full $x$-range relevant to LHC analyses. Our results are presented 
analytically for a general gauge group for detailed checks and 
validations of such future calculations.
\vspace*{0.5cm}
\end{titlepage}
%
%
Fully consistent analyses of hard processes with initial-state 
hadrons at the $($next-to$)^{n}$-leading order (N$^{\:\!n\:\!}$LO) 
of renormalization-group improved perturbative QCD require parton 
distributions functions (PDFs) evolved with the $(n\!+\!1)$-loop 
splitting functions.  
Over the past years, N$^2$LO ($=$ NNLO) has become the standard 
approximation for many processes. Following pioneering computations
of their lowest integer-$N$ Mellin moments in 
refs.~\cite{Mom3loop1,Mom3loop2}, the corresponding 3-loop splitting
functions were obtained in refs.~\cite{mvvPns,mvvPsg}.

For certain benchmark cases, in particular Higgs-boson production at
the LHC
\cite{Higgs},
N$^2$LO calculations are not sufficiently accurate, 
hence the 4-loop splitting functions need to be calculated. These
have been determined for the flavour non-singlet quark-quark case
in ref.~\cite{MRUVV1} -- analytically in the limit of a large number 
of colours $\nc$, and numerically for the remaining contributions --
and for the (next-to-)$\,$leading contributions for a large number 
of flavours $\nf$ in ref.~\cite{DRUVV}. 

Here we present, as a first significant step towards at least 
approximate expressions for the 4-loop singlet splitting functions
for use in phenomenological analyses, their lowest four even moments 
$N = 2\,\dots,\,8$ in the standard \MSb\ scheme, thus extending the 
computations of ref.~\cite{Mom3loop2} by one order in the strong 
coupling $\als$.
Following the approach of refs.~\cite{Mom3loop2,mvvPsg} our 
calculations are performed via physical quantities in deep-inelastic 
scattering, i.e., instead of working with 4-loop off-shell 
flavour-singlet operator matrix elements (OMEs) which, at this point, 
is still theoretically challenging. 
Our present results, obtained analytically for a general gauge group, 
should also be useful for checking and validating future OME 
computations of these quantities.

The evolution equations for the flavour-singlet quark and gluon PDFs 
of hadrons,
\beq
\label{sgPDFs}
  q_{\sf s}^{}(x,\mus) \; = \; 
  \sum_{i=1}^{n_{\!f}} \left[\,
     q_i^{}(x,\mus) + \,\bar{q}_i^{}(x,\mus) 
  \:\!\right]
\quad \mbox{and} \quad g(x,\mus)
\; ,
\eeq
are
\beq
\label{sgEvol}
  \frac{d}{d \ln\mus} \;
  \Big( \begin{array}{c}
          \! q_{\sf s}^{} \!\! \\ \!g\!  
        \end{array} 
  \Big)
  \: = \: \left(
    \begin{array}{cc} \! P_{\rm qq} & P_{\rm qg} \!\!\! \\
                      \! P_{\rm gq} & P_{\rm gg} \!\!\! \end{array} 
          \right)
  \otimes
  \Big( \begin{array}{c} 
          \!q_{\rm s}^{}\!\! \\ \!g\!  
        \end{array} 
  \Big)
  \:\: .
\eeq
Here $\otimes$ represents the Mellin convolution in the momentum 
variable, and $\mu_f^{}$ is the factorization scale. 
For the determination of the splitting functions $P_{\,\rm ik}$, 
the renormalization scale can be identified with $\mu_f^{}$ without 
loss of information.
The even-$N$ moments of splitting functions in eq.~(\ref{sgEvol}) 
are identical to the anomalous dimensions of twist-2 spin-$N$ 
operators up to a conventional sign,
\beq
\label{gamP}
  \gamma_{\,\rm ik}^{}(N,\als) \; = \;
  - \int_0^1 \!dx\:\, x^{\,N-1}\, P_{\,\rm ik}^{}(x,\als)
\:\: .
\eeq
Their perturbative expansions can be written as
\beq
\label{gamexp}
  \gamma_{\,\rm ik}^{}\left(N,\als\right) \; = \; \sum_{n=0}\,
  \ar(n+1) \,\gamma^{\,(n)}_{\,\rm ik}(x)
\quad \mbox{with} \quad \ars \:\equiv\; \frct{\als(\mus)}{4\pi}
  \:\: .
\eeq
 
The quark-quark entry in eq.~(\ref{gamP}) can be expressed as
$\gamma_{\,\rm qq} = \gamma_{\,\rm ns}^{\,+} + \gamma_{\,\rm ps}^{}$
in terms of the non-singlet anomalous dimension 
$\gamma_{\,\rm ns}^{\,+}$ for quark-antiquark sums addressed at 
four loops in ref.~\cite{MRUVV1} and a pure-singlet contribution 
$\gamma_{\,\rm ps}^{}$ which is suppressed at $N \gg 1$.
At asymptotically large $N$ the diagonal \MSb\ entries 
$\gamma_{\,\rm kk}(N)$ in eq.~(\ref{sgEvol}) are governed by the 
(lightlike) cusp anomalous dimensions $A_{\rm k}$ \cite{Korch89}, viz 
$ \gamma_{\,\rm kk}(N) \,=\, A_{\rm k} \ln\, N + {\cal O}(1) $,
which are now fully known at four loops 
\cite{Henn:2019swt,vonManteuffel:2020vjv}.

The 4-loop contributions to the pure-singlet anomalous dimensions
in eq.~(\ref{gamexp}) at $N=2,\,4,\,6$ are
\bea
\label{eq:GpsN2}
 \lefteqn{ \hspn \gamma_{\,\rm ps}^{\,(3)}(N\!=\!2) \:\equal\:
    \nf\, \* \cft \, \* \left( 
      { 227938 \over 2187 }
    + { 1952 \over 81 }\, \* \z3
    + { 256 \over 9 }\, \* \z4
    - { 640 \over 3 }\, \* \z5 \right)
} \nn \\[0.5mm] &&  \mbox{\hspn} 
  + \,\nf\, \* \ca \* \cfs \, \* \left(
    - { 162658 \over 6561 }
    + { 8048 \over 27 }\, \* \z3
    - { 1664 \over 9 }\, \* \z4
    + { 320 \over 9 }\, \* \z5 \right)
\nn \\[0.5mm] && \mbox{\hspn}
  + \,\nf\, \* \cas\, \* \cf \, \*  \left(
    - { 410299 \over 6561 }
    - { 26896 \over 81 }\, \* \z3
    + { 1408 \over 9 }\, \* \z4
    + { 4480 \over 27 }\, \* \z5 \right)
\nn \\[0.5mm] && \mbox{\hspn}
  + \,\nf \, \* \dfRRnr \, \* \left( 
      { 1024 \over 9 }
    + { 256 \over 9 }\, \* \z3
    - { 2560 \over 9 }\, \* \z5 \right)
  - \,\nfs\, \* \cfs \, \*  \left(
      { 73772 \over 6561 }
    + { 5248 \over 81 }\, \* \z3
    - { 320 \over 9 }\, \* \z4 \right)
\nn \\[0.5mm] && \mbox{\hspn}
  + \,\nfs\, \* \ca \* \cf \, \* \left( 
      { 160648 \over 6561 } + 48\, \* \z3
    - { 320 \over 9 }\, \* \z4 \right)
  + \nft\, \* \cf \, \*  \left(
    - { 1712 \over 729 }
    + { 128 \over 27 }\, \* \z3 \right)
\; , \\[2mm]
\label{eq:GpsN4}
 \lefteqn{ \hspn \gamma_{\,\rm ps}^{\,(3)}(N\!=\!4) \:\equal\:
    \nf\, \* \cft \, \* \left( 
      { 1995890620891 \over 52488000000 }
    - { 897403 \over 202500 }\, \* \z3
    + { 18997 \over 2250 }\, \* \z4
    - { 484 \over 15 }\, \* \z5 \right)
} \nn \\[0.5mm] &&  \mbox{\hspn}
  + \,\nf\, \* \ca\, \* \cfs \, \* \left( 
      { 209865827521 \over 26244000000 }
    + { 6743539 \over 202500 }\, \* \z3
    - { 29161 \over 750 }\, \* \z4
    + { 242 \over 45 }\, \* \z5 \right)
\nn \\[0.5mm] && \mbox{\hspn}
  + \,\nf\, \* \cas\, \* \cf \, \*  \left(
    - { 55187654921 \over 3280500000 }
    - { 3104267 \over 67500 }\, \* \z3
    + { 34243 \over 1125 }\, \* \z4
    + { 3164 \over 135 }\, \* \z5 \right)
\nn \\[0.5mm] && \mbox{\hspn}
  + \,\nf\, \* \dfRRnr \, \* \left( 
      { 172231 \over 675 }
    - { 5368 \over 25 }\, \* \z3
    - { 3728 \over 45 }\, \* \z5 \right)
  - \,\nfs\, \* \cfs \, \*  \left(
      { 141522185707 \over 26244000000 }
\right.\nn \\[0.5mm] && \mbox{} \left.
    + { 1207 \over 135 }\, \* \z3
    - { 242 \over 45 }\, \* \z4 \right)
  + \,\nfs\, \* \ca\, \* \cf \, \* \left( 
      { 9398360351 \over 1640250000 }
    + { 57877 \over 10125 }\, \* \z3
    - { 242 \over 45 }\, \* \z4 \right)
\nn \\[0.5mm] && \mbox{\hspn}
  + \,\nft\, \* \cf \, \*  \left(
    - { 46099151 \over 72900000 }
    + { 484 \over 675 }\, \* \z3 \right)
\; , \\[2mm]
\label{eq:GpsN6}
 \lefteqn{ \hspn \gamma_{\,\rm ps}^{\,(3)}(N\!=\!6) \:\equal\:
    \nf\, \* \cft \, \* \left( 
      { 140565274663259489 \over 5403265623000000 }
    - { 62727544 \over 24310125 }\, \* \z3
    + { 343156 \over 77175 }\, \* \z4
    - { 1936 \over 147 }\, \* \z5 \right)
} \nn \\[1mm] &&  \mbox{\hspn}
  + \,\nf\, \* \ca\, \* \cfs \, \* \left( 
      { 336481838777617 \over 360217708200000 }
    + { 2111992 \over 324135 }\, \* \z3
    - { 1389806 \over 77175 }\, \* \z4
    + { 968 \over 441 }\, \* \z5 \right)
\nn \\[1mm] && \mbox{\hspn}
  + \,\nf\, \* \cas\, \* \cf \, \*  \left(
    - { 6194882229735067 \over 864522499680000 }
    - { 2396237 \over 165375 }\, \* \z3
    + { 41866 \over 3087 }\, \* \z4
    + { 9544 \over 1323 }\, \* \z5 \right)
\nn \\[1mm] && \mbox{\hspn}
  + \,\nf\, \* \dfRRnr \, \* \left( 
      { 64697569 \over 330750 }
    - { 426976 \over 3675 }\, \* \z3
    - { 39808 \over 441 }\, \* \z5 \right)
\nn \\[1mm] && \mbox{\hspn}
  - \,\nfs\, \* \cfs \, \*  \left(
      { 812984663253277 \over 270163281150000 }
    + { 2594876 \over 694575 }\, \* \z3
    - { 968 \over 441 }\, \* \z4 \right)
  + \,\nfs\, \* \ca\, \* \cf \, \* \left( 
      { 3092531515013 \over 964868861250 }
\right.\nn \\[1mm] && \mbox{} \left.
    + { 217432 \over 99225 }\, \* \z3
    - { 968 \over 441 }\, \* \z4 \right)
  + \,\nft\, \* \cf \, \*  \left(
    - { 19597073837 \over 61261515000 }
    + { 1936 \over 6615 }\, \* \z3 \right)
\; .
\eea
The complete $qq$ entries are obtained by adding the non-singlet 
contributions in app.~B of ref.~\cite{MRUVV1}.

The corresponding results for the off-diagonal splitting functions 
are given by
\bea
\label{eq:GqgN2}
 \lefteqn{ \hspn \gamma_{\,\rm qg}^{\,(3)}(N\!=\!2) \:\equal\:
    \nf\, \* \cft \, \* \left( 
      { 16489 \over 729 }
    + { 736 \over 81 }\, \* \z3
    + { 256 \over 9 }\, \* \z4
    - { 320 \over 3 }\, \* \z5 \right)
} \nn \\[0.5mm] &&  \mbox{\hspn}
  + \,\nf\, \* \cat \, \*  \left(
    - { 88769 \over 729 }
    + { 31112 \over 81 }\, \* \z3 - 132\, \* \z4
    - { 3560 \over 27 }\, \* \z5 \right)
  - \,\nf\, \* \ca \* \cfs \, \*  \left(
      { 1153727 \over 13122 }
    - { 7108 \over 81 }\, \* \z3
\right. \nn \\[0.5mm] &&  \mbox{} \left.
    + { 1136 \over 9 }\, \* \z4
    - { 2000 \over 9 }\, \* \z5 \right)
  + \,\nf\, \* \cas\, \* \cf \, \* \left( 
      { 763868 \over 6561 }
    - { 12808 \over 27 }\, \* \z3
    + { 2068 \over 9 }\, \* \z4
    + { 40 \over 9 }\, \* \z5 \right)
\nn \\[0.5mm] && \mbox{\hspn}
  + \,\nf \,\, \* \dfRAna \, \* \left( 
      { 368 \over 9 }
    - { 992 \over 9 }\, \* \z3
    - { 2560 \over 9 }\, \* \z5 \right)
  - \,\nfs\, \* \cfs \, \*  \left(
      { 110714 \over 6561 }
    + { 272 \over 9 }\, \* \z3
    - { 224 \over 9 }\, \* \z4 \right)
\nn \\[0.5mm] && \mbox{\hspn}
  + \,\nfs\, \* \ca\, \* \cf \, \* \left( 
      { 249310 \over 6561 }
    + { 5632 \over 81 }\, \* \z3
    - { 440 \over 9 }\, \* \z4 \right)
  + \,\nfs\, \* \cas \, \* \left( 
      { 48625 \over 2187 }
    - { 3572 \over 81 }\, \* \z3 
\right. \nn \\[0.5mm] &&  \mbox{} \left.
    + 24\, \* \z4
    + { 160 \over 27 }\, \* \z5 \right)
  + \,\nfs \,\, \* \dfRRna\, \*  \left(
    - { 928 \over 9 }
    - { 640 \over 9 }\, \* \z3
    + { 2560 \over 9 }\, \* \z5 \right)
\nn \\[0.5mm] && \mbox{\hspn}
  + \,\nft\, \* \cf \, \*  \left(
    - { 8744 \over 2187 }
    + { 128 \over 27 }\, \* \z3 \right)
  + \,\nft\, \* \ca \, \* \left( 
      { 3385 \over 2187 }
    - { 176 \over 81 }\, \* \z3 \right)
\; , \\[2mm]
\label{eq:GqgN4}
 \lefteqn{ \hspn \gamma_{\,\rm qg}^{\,(3)}(N\!=\!4) \:\equal\:
    \nf\, \* \cft \, \*  \left(
    - { 8103828487201 \over 104976000000 }
    + { 5100751 \over 81000 }\, \* \z3
    + { 154589 \over 4500 }\, \* \z4
    - { 3158 \over 45 }\, \* \z5 \right)
} \nn \\[0.5mm] &&  \mbox{\hspn}
  + \,\nf\, \* \ca\, \* \cfs \, \* \left( 
      { 5121012352507 \over 26244000000 }
    - { 48971263 \over 405000 }\, \* \z3
    - { 143489 \over 750 }\, \* \z4
    + { 951 \over 5 }\, \* \z5 \right)
\nn \\[0.5mm] && \mbox{\hspn}
  + \,\nf\, \* \cas\, \* \cf \, \*  \left(
    - { 314624947013 \over 1312200000 }
    - { 2024593 \over 9000 }\, \* \z3
    + { 1674889 \over 4500 }\, \* \z4
    + { 1237 \over 45 }\, \* \z5 \right)
\nn \\[0.5mm] && \mbox{\hspn}
  + \,\nf\, \* \cat \, \* \left( 
      { 143199094853 \over 1458000000 }
    + { 11938031 \over 45000 }\, \* \z3
    - { 26904 \over 125 }\, \* \z4
    - { 17917 \over 135 }\, \* \z5 \right)
\nn \\[0.5mm] && \mbox{\hspn}
  + \,\nf\, \* \dfRAna \, \*  \left(
    - { 12196 \over 135 }
    - { 81008 \over 225 }\, \* \z3
    + { 15976 \over 45 }\, \* \z5 \right)
\nn \\[0.5mm] && \mbox{\hspn}
  + \,\nfs\, \* \cfs \, \* \left( 
      { 37295583467 \over 26244000000 }
    - { 1400864 \over 50625 }\, \* \z3
    + { 707 \over 45 }\, \* \z4 \right)
  + \,\nfs\, \* \ca\, \* \cf \, \* \left( 
      { 217239001681 \over 13122000000 }
\right.\nn \\[0.5mm] && \mbox{} \left.
    + { 4497112 \over 50625 }\, \* \z3
    - { 103669 \over 2250 }\, \* \z4 \right)
  + \,\nfs\, \* \cas \, \*  \left(
    - { 7131194093 \over 4374000000 }
    - { 12599759 \over 202500 }\, \* \z3
\right.\nn \\[0.5mm] && \mbox{} \left.
    + { 7591 \over 250 }\, \* \z4
    + { 664 \over 135 }\, \* \z5 \right)
  + \,\nfs\, \* \dfRRna \, \*  \left(
    - { 112424 \over 675 }
    - { 2336 \over 75 }\, \* \z3
    + { 10624 \over 45 }\, \* \z5 \right)
\nn \\[0.5mm] && \mbox{\hspn}
  + \,\nft\, \* \cf \, \*  \left(
    - { 312015851 \over 364500000 }
    + { 6644 \over 3375 }\, \* \z3 \right)
  + \,\nft\, \* \ca \, \* \left( 
      { 338346151 \over 437400000 }
    - { 5192 \over 2025 }\, \* \z3 \right)
\; , \\[2mm]
\label{eq:GqgN6}
 \lefteqn{ \hspn \gamma_{\,\rm qg}^{\,(3)}(N\!=\!6) \equal
    \,\nf\, \* \cat \, \* \left( 
      { 49981299563948069 \over 345808999872000 }
    + { 2383601783 \over 12965400 }\, \* \z3
    - { 689907 \over 3430 }\, \* \z4
    - { 159724 \over 1323 }\, \* \z5 \right)
} \nn \\[0.5mm] &&  \mbox{\hspn}
  + \,\nf\, \* \ca\, \* \cfs \, \* \left( 
      { 324177529264517279 \over 960580555200000 }
    - { 1154450237 \over 9724050 }\, \* \z3
    - { 28952417 \over 154350 }\, \* \z4
    + { 9832 \over 441 }\, \* \z5 \right)
\nn \\[0.5mm] && \mbox{\hspn}
  + \,\nf\, \* \cas\, \* \cf \, \*  \left(
    - { 627686002393628869 \over 1729044999360000 }
    - { 6170262713 \over 48620250 }\, \* \z3
    + { 1096679 \over 3087 }\, \* \z4
    + { 47774 \over 441 }\, \* \z5 \right)
\nn \\[0.5mm] && \mbox{\hspn}
  + \nf\, \* \cft \, \*  \left(
    - { 2912197809548779709 \over 21613062492000000 }
    + { 1026604067 \over 24310125 }\, \* \z3
    + { 2582141 \over 77175 }\, \* \z4
    + { 1328 \over 147 }\, \* \z5 \right)
\nn \\[0.5mm] && \mbox{\hspn}
  + \,\nf\, \* \dfRAna \, \*  \left(
    - { 23820479 \over 264600 }
    - { 11627738 \over 33075 }\, \* \z3
    + { 28624 \over 63 }\, \* \z5 \right)
\nn \\[0.5mm] && \mbox{\hspn}
  + \,\nfs\, \* \cfs \, \* \left(
      { 1942638296203817 \over 540326562300000 }
    - { 113578219 \over 4862025 }\, \* \z3
    + { 28724 \over 2205 }\, \* \z4 \right) 
\nn \\[0.5mm] && \mbox{\hspn}
  + \,\nfs\, \* \ca\, \* \cf \, \* \left( 
      { 3261418656515051 \over 216130624920000 }
    + { 122909317 \over 1620675 }\, \* \z3
    - { 600626 \over 15435 }\, \* \z4 \right)
\nn \\[0.5mm] && \mbox{\hspn}
  + \,\nfs\, \* \cas \, \*  \left(
    - { 55264268415947 \over 6175160712000 }
    - { 38177677 \over 720300 }\, \* \z3
    + { 133186 \over 5145 }\, \* \z4
    + { 5360 \over 1323 }\, \* \z5 \right)
\nn \\[0.5mm] && \mbox{\hspn}
  + \,\nfs\, \* \dfRRna \, \*  \left(
    - { 665983 \over 4725 }
    - { 192736 \over 6615 }\, \* \z3
    + { 85760 \over 441 }\, \* \z5 \right)
\nn \\[0.5mm] && \mbox{\hspn}
  + \,\nft\, \* \cf \, \*  \left(
    - { 1262351231147 \over 2572983630000 }
    + { 15268 \over 9261 }\, \* \z3 \right)
  + \,\nft\, \* \ca \, \* \left(
      { 34431246007 \over 55135363500 }
    - { 8866 \over 3969 }\, \* \z3 \right) 
\; .
\eea
and
\bea
\label{eq:GgqN2}
 \lefteqn{ \hspn \gamma_{\,\rm gq}^{\,(3)}(N\!=\!2) \:\equal\:
 \,-\, \gamma_{\,\rm qq}^{\,(3)}(N\!=\!2) 
 \; , } \\[2mm]
\label{eq:GgqN4}
 \lefteqn{ \hspn \gamma_{\,\rm gq}^{\,(3)}(N\!=\!4) \:\equal\:
    \cff \, \*  \left(
    - { 1438431824489 \over 17496000000 }
    - { 21061493 \over 101250 }\, \* \z3
    + { 259 \over 5 }\, \* \z4
    + { 14408 \over 45 }\, \* \z5 \right)
} \nn \\[0.5mm] &&  \mbox{\hspn}
  + \,\ca\, \* \cft \, \* \left( 
      { 270563159561 \over 8748000000 }
    + { 6105179 \over 101250 }\, \* \z3
    + { 5917 \over 750 }\, \* \z4
    - { 17488 \over 45 }\, \* \z5 \right)
\nn \\[0.5mm] && \mbox{\hspn}
  + \,\cas\, \* \cfs \, \* \left( 
      { 1259255579057 \over 4374000000 }
    + { 16267093 \over 67500 }\, \* \z3
    - { 25621 \over 250 }\, \* \z4
    + { 1484 \over 45 }\, \* \z5 \right)
\nn \\[0.5mm] && \mbox{\hspn}
  + \,\cat\, \* \cf \, \*  \left(
    - { 632341192829 \over 2187000000 }
    - { 1120409 \over 8100 }\, \* \z3
    + { 16048 \over 375 }\, \* \z4
    + { 8782 \over 135 }\, \* \z5 \right)
\nn \\[0.5mm] && \mbox{\hspn}
  + \,\dfRAnr \, \*  \left(
    - { 12196 \over 135 }
    - { 81008 \over 225 }\, \* \z3
    + { 15976 \over 45 }\, \* \z5 \right)
  + \,\nf\, \* \cft \, \*  \left(
    - { 316818132031 \over 3280500000 }
\right.\nn \\[0.5mm] && \mbox{} \left.
    + { 411629 \over 16875 }\, \* \z3
    - { 4582 \over 225 }\, \* \z4
    + { 352 \over 3 }\, \* \z5 \right)
  + \,\nf\, \* \ca\, \* \cfs \, \* \left( 
      { 569679966383 \over 6561000000 }
    - { 13919446 \over 50625 }\, \* \z3
\right.\nn \\[0.5mm] && \mbox{} \left.
    + { 12501 \over 125 }\, \* \z4
    - { 176 \over 9 }\, \* \z5 \right)
  + \nf\, \* \dfRRnr \, \*  \left(
    - { 112424 \over 675 }
    - { 2336 \over 75 }\, \* \z3
    + { 10624 \over 45 }\, \* \z5 \right)
\nn \\[0.5mm] && \mbox{\hspn}
  + \,\nf\, \* \cas\, \* \cf \, \* \left( 
      { 2203719743 \over 52488000 }
    + { 2857549 \over 11250 }\, \* \z3
    - { 89599 \over 1125 }\, \* \z4
    - { 11872 \over 135 }\, \* \z5 \right)
\nn \\[0.5mm] && \mbox{\hspn}
  + \,\nfs\, \* \cfs \, \* \left( 
      { 9798304643 \over 3280500000 }
    + { 17096 \over 675 }\, \* \z3
    - { 704 \over 45 }\, \* \z4 \right)
  + \,\nfs\, \* \ca\, \* \cf \, \*  \left(
    - { 1608863899 \over 328050000 }
\right.\nn \\[0.5mm] && \mbox{} \left.
    - { 39416 \over 2025 }\, \* \z3
    + { 704 \over 45 }\, \* \z4 \right)
  + \,\nft\, \* \cf \, \* \left( 
      { 3990397 \over 2733750 }
    - { 704 \over 405 }\, \* \z3 \right)
 \; , \\[2mm]
\label{eq:GgqN6}
 \lefteqn{ \hspn \gamma_{\,\rm gq}^{\,(3)}(N\!=\!6) \:\equal\:
    \cff \, \*  \left(
    - { 27548846012571077 \over 225136067625000 }
    - { 28516720088 \over 121550625 }\, \* \z3
    + { 6416 \over 105 }\, \* \z4
    + { 260192 \over 735 }\, \* \z5 \right)
} \nn \\[0.5mm] &&  \mbox{\hspn}
  + \,\ca\, \* \cft \, \* \left( 
      { 15370144370986843 \over 90054427050000 }
    + { 23472335174 \over 121550625 }\, \* \z3
    - { 1023364 \over 25725 }\, \* \z4
    - { 370016 \over 735 }\, \* \z5 \right)
\nn \\[0.5mm] && \mbox{\hspn}
  + \,\cas\, \* \cfs \, \* \left( 
      { 58564721355491371 \over 720435416400000 }
    + { 10781187328 \over 121550625 }\, \* \z3
    - { 1215814 \over 25725 }\, \* \z4
    + { 373832 \over 2205 }\, \* \z5 \right)
\nn \\[0.5mm] && \mbox{\hspn}
  + \,\cat\, \* \cf \, \*  \left(
    - { 133292466369681947 \over 864522499680000 }
    - { 226736591 \over 2701125 }\, \* \z3
    + { 667258 \over 25725 }\, \* \z4
    + { 67288 \over 6615 }\, \* \z5 \right)
\nn \\[0.5mm] && \mbox{\hspn}
  + \,\dfRAnr \, \*  \left(
    - { 23820479 \over 330750 }
    - { 46510952 \over 165375 }\, \* \z3
    + { 114496 \over 315 }\, \* \z5 \right)
\nn \\[0.5mm] && \mbox{\hspn}
  + \,\nf\, \* \cft \, \*  \left(
    - { 75665018489451691 \over 1350816405750000 }
    + { 187225352 \over 24310125 }\, \* \z3
    - { 36352 \over 2205 }\, \* \z4
    + { 1408 \over 21 }\, \* \z5 \right)
\nn \\[0.5mm] && \mbox{\hspn}
  + \,\nf\, \* \ca\, \* \cfs \, \* \left( 
      { 331099053590779 \over 6003628470000 }
    - { 3771301108 \over 24310125 }\, \* \z3
    + { 4877248 \over 77175 }\, \* \z4
    - { 704 \over 63 }\, \* \z5 \right)
\nn \\[0.5mm] && \mbox{\hspn}
  + \,\nf\, \* \cas\, \* \cf \, \* \left( 
      { 40511222207957 \over 3430644840000 }
    + { 3610221368 \over 24310125 }\, \* \z3
    - { 3604928 \over 77175 }\, \* \z4
    - { 65344 \over 1323 }\, \* \z5 \right)
\nn \\[0.5mm] && \mbox{\hspn}
  + \nf\, \* \dfRRnr \, \*  \left(
    - { 2663932 \over 23625 }
    - { 770944 \over 33075 }\, \* \z3
    + { 68608 \over 441 }\, \* \z5 \right)
\nn \\[0.5mm] && \mbox{\hspn}
  + \,\nfs\, \* \cfs \, \* \left(
      { 27562736653631 \over 9648688612500 }
    + { 266912 \over 19845 }\, \* \z3
    - { 2816 \over 315 }\, \* \z4 \right) 
  - \,\nfs\, \* \ca\, \* \cf \, \*  \left(
      { 301286343367 \over 110270727000 }
\right.\nn \\[0.5mm] && \mbox{} \left.
    + { 944432 \over 99225 }\, \* \z3
    - { 2816 \over 315 }\, \* \z4 \right)
  + \,\nft\, \* \cf \, \* \left( 
      { 3574461862 \over 3281866875 }
    - { 2816 \over 2835 }\, \* \z3 \right)
\; .
\eea
Finally the lowest three even moments (\ref{gamP}) of the four-loop 
gluon-gluon splitting function read
\bea
\label{eq:GggN2}
 \lefteqn{ \hspn \gamma_{\,\rm gg}^{\,(3)}(N\!=\!2) \:\equal\:
 \,-\, \gamma_{\,\rm qg}^{\,(3)}(N\!=\!2)
 \; , } \\[2mm]
\label{eq:GggN4}
 \lefteqn{ \hspn \gamma_{\,\rm gg}^{\,(3)}(N\!=\!4) \:\equal\:
    \caf \, \*
    \left( { 1502628149 \over 3375000 }
    + { 1146397 \over 11250 }\, \* \z3
    - { 504 \over 5 }\, \* \z5 \right)
  + \,\dfAAna \, \*
    \left( { 21623 \over 150 }
\right. } \nn \\[0.5mm] &&  \mbox{} \left.
    + { 15596 \over 15 }\, \* \z3
    - { 6048 \over 5 }\, \* \z5 \right)
  + \,\nf\, \* \dfRAna \, \* \left( 
      { 160091 \over 675 }
    + { 80072 \over 225 }\, \* \z3
    - { 48016 \over 45 }\, \* \z5 \right)
\nn \\[0.5mm] &&  \mbox{\hspn}
  + \,\nf\, \* \cat \, \*  \left(
    - { 20580892841 \over 72900000 }
    - { 12550223 \over 22500 }\, \* \z3
    + { 8613 \over 25 }\, \* \z4
    + { 4316 \over 27 }\, \* \z5 \right)
\nn \\[0.5mm] &&  \mbox{\hspn}
  + \nf\, \* \cas\, \* \cf \, \*  \left(
    - { 4212122951 \over 41006250 }
    + { 1170784 \over 5625 }\, \* \z3
    - { 418198 \over 1125 }\, \* \z4
    + { 17636 \over 45 }\, \* \z5 \right)
\nn \\[0.5mm] &&  \mbox{\hspn}
  + \,\nf\, \* \ca \* \cfs \, \* \left( 
      { 1913110089023 \over 26244000000 }
    + { 39313783 \over 101250 }\, \* \z3
    + { 26741 \over 750 }\, \* \z4
    - { 3082 \over 5 }\, \* \z5 \right)
\nn \\[0.5mm] &&  \mbox{\hspn}
  + \,\nf\, \* \cft \, \* \left( 
      { 34764568601 \over 2099520000 }
    - { 958343 \over 40500 }\, \* \z3
    - { 18997 \over 2250 }\, \* \z4
    + { 908 \over 45 }\, \* \z5 \right)
\nn \\[0.5mm] &&  \mbox{\hspn}
  + \,\nfs\, \* \cas \, \*  \left(
    - { 3250393649 \over 218700000 }
    + { 2969291 \over 20250 }\, \* \z3
    - { 1566 \over 25 }\, \* \z4
    - { 1276 \over 135 }\, \* \z5 \right)
\nn \\[0.5mm] &&  \mbox{\hspn}
  + \,\nfs\, \* \ca \* \cf \, \* \left( 
      { 136020246173 \over 3280500000 }
    - { 1672751 \over 10125 }\, \* \z3
    + { 15172 \over 225 }\, \* \z4 \right)
  - \,\nfs\, \* \cfs \, \*  \left(
      { 275622924731 \over 26244000000 }
\right. \nn \\[0.5mm] &&  \mbox{} \left.
    - { 253369 \over 10125 }\, \* \z3
    + { 1078 \over 225 }\, \* \z4 \right)
  + \,\nfs\,\, \* \dfRRna \, \* \left( 
      { 75788 \over 675 }
    + { 3008 \over 15 }\, \* \z3
    - { 20416 \over 45 }\, \* \z5 \right)
\nn \\[0.5mm] &&  \mbox{\hspn}
  + \,\nft\, \* \ca \, \*  \left(
    - { 20440457 \over 21870000 }
    + { 1888 \over 405 }\, \* \z3 \right)
  + \,\nft\, \* \cf \, \* \left( 
      { 1780699 \over 24300000 }
    - { 484 \over 675 }\, \* \z3 \right)
 \; , \\[2mm]
\label{eq:GggN6}
 \lefteqn{ \hspn \gamma_{\,\rm gg}^{\,(3)}(N\!=\!6) \:\equal\:
    \caf \, \* \left( 
      { 14796034088334539 \over 23053933324800 }
    + { 198201877 \over 777924 }\, \* \z3
    - { 118210 \over 441 }\, \* \z5 \right)
} \nn \\[0.5mm] &&  \mbox{\hspn}
  + \,\dfAAna \, \* \left( 
      { 1255552 \over 2205 }
    + { 2997592 \over 1323 }\, \* \z3
    - { 472840 \over 147 }\, \* \z5 \right)
\nn \\[0.5mm] &&  \mbox{\hspn}
  + \,\nf\, \* \cat \, \*  \left(
    - { 352499691830939 \over 914838624000 }
    - { 4467756563 \over 6482700 }\, \* \z3
    + { 103356 \over 245 }\, \* \z4
    + { 238544 \over 1323 }\, \* \z5 \right)
\nn \\[0.5mm] &&  \mbox{\hspn}
  + \nf\, \* \cas\, \* \cf \, \*  \left(
    - { 174297079261544753 \over 864522499680000 }
    + { 12199024283 \over 48620250 }\, \* \z3
    - { 6710594 \over 15435 }\, \* \z4
    + { 221576 \over 441 }\, \* \z5 \right)
\nn \\[0.5mm] &&  \mbox{\hspn}
  + \,\nf\, \* \ca \* \cfs \, \* \left( 
      { 51836938615212157 \over 360217708200000 }
    + { 459844342 \over 972405 }\, \* \z3
    + { 1338986 \over 77175 }\, \* \z4
    - { 334352 \over 441 }\, \* \z5 \right)
\nn \\[0.5mm] &&  \mbox{\hspn}
  + \,\nf\, \* \cft \, \* \left( 
      { 10457671535671561 \over 5403265623000000 }
    - { 100551124 \over 8103375 }\, \* \z3
    - { 343156 \over 77175 }\, \* \z4
    + { 992 \over 147 }\, \* \z5 \right)
\nn \\[0.5mm] &&  \mbox{\hspn}
  + \,\nf\, \* \dfRAna \, \* \left( 
      { 9661697 \over 22050 }
    + { 22351528 \over 33075 }\, \* \z3
    - { 726848 \over 441 }\, \* \z5 \right)
\nn \\[0.5mm] &&  \mbox{\hspn}
  + \,\nfs\, \* \cas \, \*  \left(
    - { 2273514775943 \over 294055272000 }
    + { 126516356 \over 694575 }\, \* \z3
    - { 18792 \over 245 }\, \* \z4
    - { 22000 \over 1323 }\, \* \z5 \right)
\nn \\[0.5mm] &&  \mbox{\hspn}
  + \,\nfs\, \* \ca \* \cf \, \* \left( 
      { 122395144706959 \over 2205414540000 }
    - { 133661648 \over 694575 }\, \* \z3
    + { 173704 \over 2205 }\, \* \z4 \right)
\nn \\[0.5mm] &&  \mbox{\hspn}
  + \,\nfs\, \* \cfs \, \* \left(
    - { 61017705026527 \over 5403265623000 }
    + { 2171164 \over 99225 }\, \* \z3 
    - { 4576 \over 2205 }\, \* \z4 \right) 
\nn \\[0.5mm] &&  \mbox{\hspn}
  + \,\nfs\,\, \* \dfRRna \, \* \left( 
      { 788419 \over 3675 }
    + { 180272 \over 441 }\, \* \z3
    - { 352000 \over 441 }\, \* \z5 \right)
\nn \\[0.5mm] &&  \mbox{\hspn}
  + \,\nft\, \* \ca \, \*  \left(
    - { 5226936307 \over 5250987000 }
    + { 3224 \over 567 }\, \* \z3 \right) 
  + \,\nft\, \* \cf \, \* \left(
    - { 9085701773 \over 30630757500 }
    - { 1936 \over 6615 }\, \* \z3 \right)
\; .
\eea
For brevity, we here write down the results at $N=8$ only numerically 
for the case of QCD:
\bea
  \gamma_{\,\rm ps}^{\,(3)}(N\!=\!8) & = & 
  - 24.014550\,\nf + 3.2351935\,\nfs - 0.0078892\,\nft
\;, \\
  \gamma_{\,\rm qg}^{\,(3)}(N\!=\!8) & = & \phantom{-}
    294.58768\,\nf - 135.37676\,\nfs - 3.6097756\,\nft
\;, \\
  \gamma_{\,\rm gq}^{\,(3)}(N\!=\!8) & = &
  - 2803.6441 + 436.393057\,\nf + 1.8149462\,\nfs + 0.0735886\,\nft
\;,\\
  \gamma_{\,\rm gg}^{\,(3)}(N\!=\!8) & = & \phantom{-}
    62279.744 - 17150.6967\,\nf + 785.88061\,\nfs + 1.8933103\,\nft
\;.
\eea

\pagebreak

In eqs.~(\ref{eq:GpsN2}) -- (\ref{eq:GggN6}) $\cf$ and $\ca$ are the
standard colour factors with $C_F = 4/3$ and $C_A = \nc = 3$ in QCD. 
The terms with the quartic group invariants
$d_{A}^{\,abcd}\,d_{A}^{\,abcd}$,
$d_{R}^{\,abcd}\,d_{A}^{\,abcd}$
and $d_{\,R}^{\,abcd}\,d_{\,R}^{\,abcd}$
agree with the results of 
ref.~\cite{MRUVV2} where these particular contributions were obtained 
to much higher values of $N$ using OME calculations.
All coefficients of the Riemann-$\zeta$ value $\z4 = \pi^4 / 90$ agree 
with the all-$N$ predictions in eqs.~(9) -- (12) of ref.~\cite{DV2017} 
based on the `no-$\pi^{2\:\!}$' conjecture of ref.~\cite{no-pi2}.
The $\nft$ contributions to all four anomalous dimension are known
for all $N$ \cite{DRUVV}, see also refs.~\cite{LargeNf}.

The above results lead to the numerical expansions 
\bea
\label{eq:GqqN2num}
  \gamma_{\,\rm qq}^{}(2,3) &\!=\!& 0.282942\, \als \, \left(
  1 + 0.736828\, \als + 0.517255\, \as(2) + 0.756972\, \as(3) 
  + \,\ldots \right)
\; , \nn \\
  \gamma_{\,\rm qq}^{}(2,4) &\!=\!& 0.282942\, \als \, \left(
  1 + 0.621883\, \als + 0.146133\, \as(2) + 0.362201\, \as(3) 
  + \,\ldots \right)
\; , \\[1.5mm]
\label{eq:GqqN4num}
  \gamma_{\,\rm qq}^{}(4,3) &\!=\!& 0.555274\, \als \, \left(
  1 + 0.756202\, \als + 0.672283\, \as(2) + 0.701628 \, \as(3) 
  + \,\ldots \right)
\; , \nn \\
  \gamma_{\,\rm qq}^{}(4,4) &\!=\!& 0.555274\, \als \, \left(
  1 + 0.680253\, \als + 0.427783\, \as(2) + 0.345861 \, \as(3) 
  + \,\ldots \right)
\; , \\[1.5mm]
\label{eq:GqqN6num}
  \gamma_{\,\rm qq}^{}(6,3) &\!=\!& 0.716450\, \als \, \left(
  1 + 0.725387\, \als + 0.685289\, \as(2) + 0.663440 \, \as(3) 
  + \,\ldots \right)
\; , \nn \\
  \gamma_{\,\rm qq}^{}(6,4) &\!=\!& 0.716450\, \als \, \left(
  1 + 0.648931\, \als + 0.426442\, \as(2) + 0.324781 \, \as(3) 
  + \,\ldots \right)
\; , \\[1.5mm]
\label{eq:GqqN8num}
  \gamma_{\,\rm qq}^{}(8,3) &\!=\!& 0.832237\, \als \, \left(
  1 + 0.710075\, \als + 0.650750\, \as(2) + 0.643336 \, \as(3) 
  + \,\ldots \right)
\; , \nn \\
  \gamma_{\,\rm qq}^{}(8,4) &\!=\!& 0.832237\, \als \, \left(
  1 + 0.632824\, \als + 0.423498\, \as(2) + 0.312139 \, \as(3) 
  + \,\ldots \right)
\eea
%
and
%
\bea
\label{eq:GqgN2num}
  \gamma_{\,\rm qg}^{}(2,3) &\!=\!& -0.159155\, \als \, \left(
  1 + 0.900404\, \als + 0.012215\, \as(2) - 0.055970\, \as(3) 
  + \,\ldots \right)
\; , \nn \\
  \gamma_{\,\rm qg}^{}(2,4) &\!=\!& -0.212207\, \als \, \left(
  1 + 0.900404\, \als - 0.102840\, \as(2) - 0.236731\, \as(3) 
  + \,\ldots \right)
\; , \\[1.5mm]
  \gamma_{\,\rm qg}^{}(4,3) &\!=\!& -0.087535\, \als \, \left(
  1 - 0.280121\, \als - 0.893969\, \as(2) - 0.022754\, \as(3) 
  + \,\ldots \right)
\; , \nn \\
\label{eq:GqgN4num}
  \gamma_{\,\rm qg}^{}(4,4) &\!=\!& -0.116714\, \als \, \left(
  1 - 0.280121\, \als - 0.998634\, \as(2) + 0.129659\, \as(3) 
  + \,\ldots \right)
\; , \\[1.5mm]
  \gamma_{\,\rm qg}^{}(6,3) &\!=\!& -0.062525\, \als \, \left(
  1 - 0.838938\, \als - 1.064575\, \as(2) + 0.145572\, \as(3) 
  + \,\ldots \right)
\; , \nn \\
\label{eq:GqgN6num}
  \gamma_{\,\rm qg}^{}(6,4) &\!=\!& -0.083367\, \als \, \left(
  1 - 0.838938\, \als - 1.150113\, \as(2) + 0.441744\, \as(3) 
  + \,\ldots \right)
\; , \\[1.5mm]
  \gamma_{\,\rm qg}^{}(8,3) &\!=\!& -0.049728\, \als \, \left(
  1 - 1.255845\, \als - 1.091729\, \as(2) + 0.353099\, \as(3) 
  + \,\ldots \right)
\; , \nn \\
  \gamma_{\,\rm qg}^{}(8,4) &\!=\!& -0.065430\, \als \, \left(
  1 - 1.255845\, \als - 1.160288\, \as(2) + 0.746929\, \as(3) 
  + \,\ldots \right)
\eea
for the upper row of the anomalous-dimension matrix, where the 
arguments of $\gamma_{\,ik}$ are $N$ and $\nf$; the values for 
$\nf = 5$ have been suppressed for brevity. 
The independent lower-row expansions -- the values at $N=2$ are 
fixed by the momentum sum-rule relations (\ref{eq:GgqN2}) and 
(\ref{eq:GggN2}) -- are given by
\bea
\label{eq:GgqN4num}
  \gamma_{\,\rm gq}^{}(4,3) &\!=\!& -0.077809\, \als \, \left(
  1 + 1.165483\, \als + 1.163066\, \as(2) + 1.474368\, \as(3) 
  + \,\ldots \right)
\; , \nn \\
  \gamma_{\,\rm gq}^{}(4,4) &\!=\!& -0.077809\, \als \, \left(
  1 + 1.115164\, \als + 0.823447\, \as(2) + 0.883269\, \as(3) 
  + \,\ldots \right)
\; , \\[1.5mm]
\label{eq:GgqN6num}
  \gamma_{\,\rm gq}^{}(6,3) &\!=\!& -0.044462\, \als \, \left(
  1 + 1.314556\, \als + 1.360970\, \as(2) + 1.726679\, \as(3) 
  + \,\ldots \right)
\; , \nn \\
  \gamma_{\,\rm gq}^{}(6,4) &\!=\!& -0.044462\, \als \, \left(
  1 + 1.301901\, \als + 1.051619\, \as(2) + 1.126955\, \as(3) 
  + \,\ldots \right)
\; , \\[1.5mm]
\label{eq:GgqN8num}
  \gamma_{\,\rm gq}^{}(8,3) &\!=\!& -0.031157\, \als \, \left(
  1 + 1.416509\, \als + 1.468523\, \as(2) + 1.899893\, \as(3) 
  + \,\ldots \right)
\; , \nn \\
  \gamma_{\,\rm gq}^{}(8,4) &\!=\!& -0.031157\, \als \, \left(
  1 + 1.430863\, \als + 1.183046\, \as(2) + 1.318370\, \as(3) 
  + \,\ldots \right)
\eea
and
\bea
\label{eq:GggN4num}
  \gamma_{\,\rm gg}^{}(4,3) &\!=\!& \;\; 1.161831\, \als \, \left(
  1 + 0.475446\, \als + 0.333272\, \as(2) + 0.478025\, \as(3) 
  + \,\ldots \right)
\; , \nn \\
  \gamma_{\,\rm gg}^{}(4,4) &\!=\!& \;\; 1.214882\, \als \, \left(
  1 + 0.383536\, \als + 0.121966\, \as(2) + 0.240469\, \as(3) 
  + \,\ldots \right)
\; , \\[2mm]
  \gamma_{\,\rm gg}^{}(6,3) &\!=\!& \;\; 1.574497\, \als \, \left(
  1 + 0.489287\, \als + 0.380902\, \as(2) + 0.429696\, \as(3) 
  + \,\ldots \right)
\; , \nn \\
\label{eq:GggN6num}
  \gamma_{\,\rm gg}^{}(6,4) &\!=\!& \;\; 1.627549\, \als \, \left(
  1 + 0.393705\, \als + 0.169676\, \as(2) + 0.190156\, \as(3) 
  + \,\ldots \right)
\; , \\[2mm]
  \gamma_{\,\rm gg}^{}(8,3) &\!=\!& \;\; 1.851503\, \als \, \left(
  1 + 0.497734\, \als + 0.404644\, \as(2) + 0.398779\, \as(3) 
  + \,\ldots \right)
\; , \nn \\
\label{eq:GggN8num}
  \gamma_{\,\rm gg}^{}(8,4) &\!=\!& \;\; 1.904554\, \als \, \left(
  1 + 0.401746\, \als + 0.194306\, \as(2) + 0.157133\, \as(3) 
  + \,\ldots \right)
\; .
\eea
Except for $\gamma_{\,\rm qg}^{}$ and $\gamma_{\,\rm gg}^{}$
at $N=8$ these numerical expansions have been presented before in 
ref.~\cite{avLL2018}.

The results for the $qq$ and $gg$ cases at asymptotically 
(and unphysically) large values of $N$ read
\bea
\label{eq:GqqNinfNum}
  \gamma_{\,\rm qq}^{}(N,3) &\!=\!& 
    \ars \, \gamma_{\,\rm qq}^{\,(0)}(N,3) \left(
  1 + 0.726574\, \als + 0.734054\, \as(2) + 0.664730\, \as(3) \right)
\; , \nn \\
  \gamma_{\,\rm qq}^{}(N,4) &\!=\!& 
    \ars \, \gamma_{\,\rm qq}^{\,(0)}(N,4) \left(
  1 + 0.638154\, \als + 0.509978\, \as(2) + 0.316848\, \as(3) \right)
\eea
and
\bea
\label{eq:GggNinfNum}
  \gamma_{\,\rm gg}^{}(N,3) &\!=\!& 
    \ars \, \gamma_{\,\rm gg}^{\,(0)}(N,3) \left(
  1 + 0.726574\, \als + 0.734054\, \as(2) + 0.415609 \, \as(3) \right)
\; , \nn \\
  \gamma_{\,\rm gg}^{}(N,4) &\!=\!& 
    \ars \, \gamma_{\,\rm gg}^{\,(0)}(N,4) \left(
  1 + 0.638154\, \als + 0.509978\, \as(2) + 0.064476 \, \as(3) \right)
\eea
due to their relation to the cusp anomalous dimensions $A_{\rm k}$.
The quark and gluon results are identical up to the `Casimir scaling'
of the prefactors, 
$ \gamma_{\,\rm qq}^{\,(0)}(N,\nf) = 4\,C_F $
and
$ \gamma_{\,\rm gg}^{\,(0)}(N,\nf) = 4\,C_A $, to three loops
and are related by a generalized (not numerical, except in the 
large-$\nc$ limit, due to the presence of the quartic group invariants) 
Casimir scaling \cite{Dixon17,MRUVV2} at four loops.

The relative size of the N$^2$LO and N$^3$LO contributions in
eqs.~(\ref{eq:GqqN2num}) -- (\ref{eq:GggNinfNum}) is illustrated in 
fig.~1 for $\nf = 4$ at $\als = 0.2$: 
The N$^3$LO corrections amount to less than 1\%, and less than 0.5\% of 
the NLO results except for $P_{\rm gq}^{}$, the quantity with the lowest 
leading-order values, at $N \geq 4$. 
Unlike in the quark case, see also ref.~\cite{Herzog:2018kwj} where 
also a first estimate of the five-loop contribution to $A_{\rm q}$ has 
been obtained, the N$^2$LO and N$^3$LO large-$N$ limits 
in the gluon case do not, in general, roughly agree with values in the 
range $4 \leq N \leq 8$ normalized as in eqs.~(\ref{eq:GqqN2num}) 
-- (\ref{eq:GggN8num}).

The resulting low-$N$ expansion for the evolution (\ref{sgEvol}) of the 
singlet quark and gluon PDFs is illustrated in fig.~2 for the schematic
but sufficiently realistic order-independent model input \cite{mvvPsg}
\bea
\label{qsgInp}
  xq_{\sf s}^{}(x,\mu_{0}^{\,2}) & =  &
  0.6\: x^{\, -0.3} (1-x)^{3.5}\, \left(1 + 5.0\: x^{\, 0.8\,} \right)
\: , \nn \\[-0.5mm]
  x\:\!g (x,\mu_{0}^{\,2})\: & =  &
  1.6\: x^{\, -0.3} (1-x)^{4.5}\, \left(1 - 0.6\: x^{\, 0.3\,} \right)
\eea
with $\als(\mu_{0}^{\,2}) = 0.2$ and $\nf=4$.
The N$^3$LO corrections are very small at the standard choice
$\mu_{r}^{}= \mu_{f}^{}\equiv \mu_{0}^{}$ of the renormalization scale.
They lead to a reduction of the scale dependence to about 1\% (full width) 
at $N \geq 4$ for the conventional range \mbox{$\frac{1}{4}\: 
\mu_{f}^{\,2} \,\leq\, \mu_{r}^{\,2} \,\leq 4\: \mu_{f}^{\,2}$}.

\begin{figure}[p]
\centerline{\hspace*{-2mm}\epsfig{file=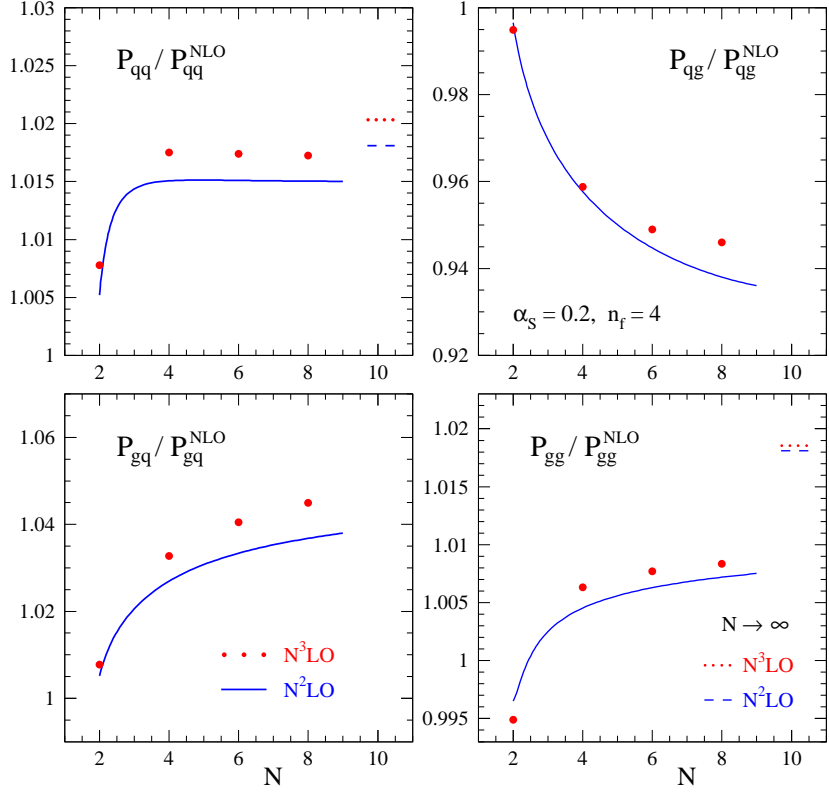,width=11cm,angle=0}}
\vspace{-3mm}
\caption{\label{Fig1}
Moments of the splitting functions (\ref{sgEvol}) at NNLO (lines) and 
N$^3$LO (even-$N$ points) at $\als = 0.2$ and $\nf = 4$, normalized to the 
NLO results. Also shown are the $qq$ and $gg$ large-$N$ limits.
}
\end{figure}
\begin{figure}[p]
\centerline{\hspace*{-2mm}\epsfig{file=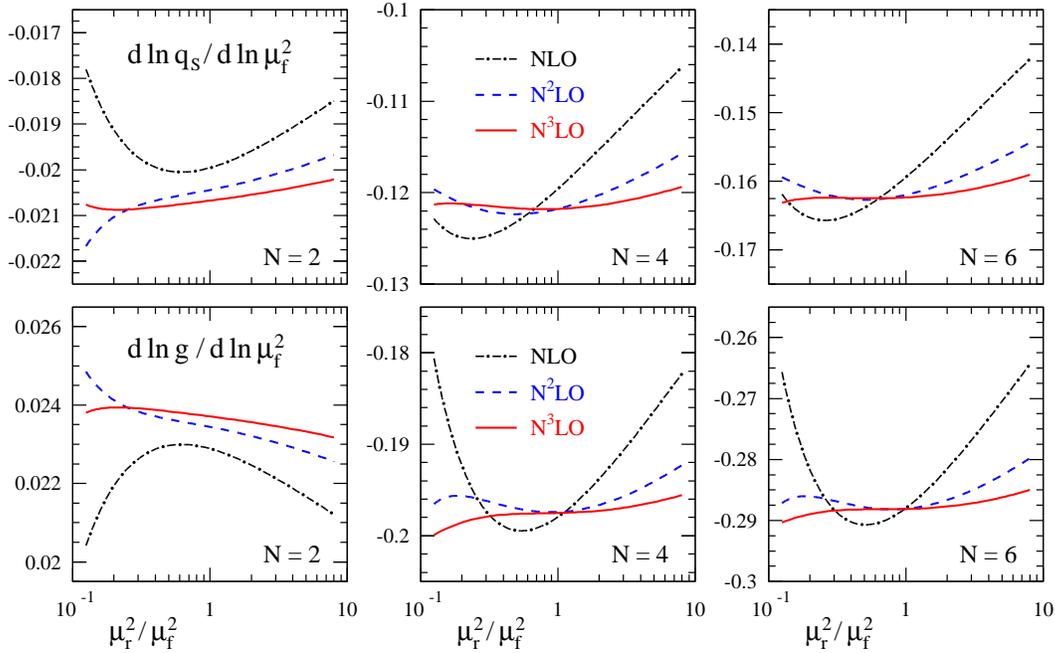,width=14cm,angle=0}}
\vspace{-3mm}
\caption{\label{Fig2}
The dependence of the logarithmic factorization-scale derivatives of
the singlet PDFs on the renormalization scale $\mu_{r}^{}$ at $N=2$
(where the very small scaling violations of $q_{\sf s}^{}$ and $g$ are
related by the momentum sum rule), $N=4$ and $N=6$ for the initial 
distributions (\protect\ref{qsgInp}).
}
\end{figure}

To summarize, we have employed the theoretical framework of 
refs.~\cite{Mom3loop1,Mom3loop2,mvvPns,mvvPsg} together with an 
optimized in-house version of the FORM \cite{FORM4} program {\sc Forcer} 
for 4-loop propagator integrals \cite{Forcer} to compute the moments
$N=2,\,4,\,6$, and 8 of all N$^3$LO flavour-singlet splitting functions.
The numerical effect of these contributions is small, but more work is
needed to arrive at sufficient `data' for a N$^3$LO analogue of the 
earlier approximate N$^2$LO splitting functions of ref.~\cite{NVappr}.

%
\subsection*{Acknowledgements}

\vspace*{-4mm}
\noindent 
This work has been supported 
by the {\it European Research Council}$\,$ (ERC) via the grants
320651 ({\it HEPGAME}) and 694712 ({\it PertQCD}), 
by the {\it European Cooperation in Science and Technology (COST)} via
{\it COST Action CA16201 PARTICLEFACE},
by the {\it Deutsche Forschungsgemeinschaft} (DFG) through the Research
Unit FOR 2926, {\it Next Generation pQCD for Hadron Structure: Preparing 
for the EIC}, project number 40824754 and DFG grant MO~1801/4-1,
by the {\it Swiss National Science Foundation} (SNSF) grant 179016,
by the {\it JSPS KAKENHI} grants 19K03831 and 21K03583,
and by the UK {\it Science \& Technology Facilities Council}$\,$ (STFC) 
grants ST/L000431/1 and ST/T000988/1.
Some of our computations were carried out on the Dutch national 
e-infrastruc\-ture with the support of the SURF Cooperative and the 
PDP Group at Nikhef, and on the {\tt ulgqcd} computer cluster in 
Liverpool which was funded by the STFC grant ST/H008837/1. 
 
{\small
\addtolength{\baselineskip}{-2mm}

}

\end{document}